\documentclass[twocolumn,a4paper,amsmath,amssymb,showpacs,pre]{revtex4}

\usepackage{graphicx}
\usepackage{epstopdf}
\usepackage{epsfig}
\usepackage{dcolumn}
\usepackage{bm}
\usepackage{ushort}
\usepackage{color}
\usepackage{psfrag}

\newcommand{\e}{{\rm e}}
\newcommand{\rmi}{{\rm i}}
\renewcommand{\d}{{\rm d}}
\newcommand{\p}{{ p}}
\newcommand{\q}{{ q}}
\newcommand{\s}{{ s}}
\newcommand{\x}{{ x}}
\newcommand{\xx}{{ x}}
\newcommand{\pp}{{ p}}

\begin{document}

\title{A Phase-Space Approach for Propagating Field-Field
  Correlation Functions}
\author{Gabriele Gradoni} 
\author{Stephen C. Creagh}%
\author{Gregor Tanner}%
\affiliation{%
 School of Mathematical Sciences, University of Nottingham, University Park, NG7 2RD, United Kingdom
 }%
\author{Christopher Smartt}
\author{David Thomas}
\affiliation{%
 George Green Institute for Electromagnetics Research, University of Nottingham, University Park, NG7 2RD, United Kingdom
 }%

\date{\today}

\begin{abstract}
We show that radiation from complex and inherently random but correlated
wave sources can be modelled efficiently by using an 
approach based on the Wigner distribution function. 
Our method exploits the connection
between correlation functions and the Wigner function and
admits in its simplest approximation a direct representation in terms of
the evolution of ray densities in phase space. 
We show that next leading order corrections to the ray-tracing approximation lead to
Airy-function type phase space propagators.
By exploiting the exact Wigner function propagator, inherently wave-like 
effects such as evanescent decay or radiation from more 
heterogeneous sources as well as diffraction and reflections 
can be included and analysed.
We discuss in particular the role of evanescent waves in the near-field of non-paraxial sources and give explicit expressions for the growth rate of the correlation length as function of the distance from the source.
Furthermore, results for the reflection of partially coherent sources from flat mirrors are given.
 We focus here on electromagnetic sources at microwave frequencies and modelling efforts in the context of  electromagnetic compatibility. 
%
%
\end{abstract}

\pacs{41.20.Jb,   
03.65.Sq, 	
42.30.Kq,   
42.15.Dp 	
}
\maketitle

\section{Introduction}
Predicting the properties of wave fields in complex environments
is an extremely challenging task of crucial importance 
to a wide variety of technological and engineering applications,
such as vibroacoustics  \cite{Tanner2014547} or electromagnetic (EM) 
wave modelling.  In particular, characterising the radiation of EM sources reliably, 
both in free space and within enclosures, is a longstanding research issue. 
In the context of electromagnetic compatibility (EMC), digital circuits and large printed 
circuit boards (PCB) embed thousands of electronic devices and
metallic tracks and can produce fields reaching dangerous but 
hard-to-predict levels \cite{montrose2004}. 

In this paper, we set out an approach for propagating such complex
and statistically characterized wave fields exploiting  Wigner distribution function (WDF) techniques. 
This approach has its origin in quantum mechanics 
\cite{Hillery1984}, but has more recently found widespread 
attention in optics, see  \cite{Dragoman_1997,torre2005,Alonso_2011}
for an overview. The WDF formalism offers a direct route to pure ray-tracing 
approximations in an operator implementation \cite{Tanner2014547}, 
while still capturing in its exact formulation the full wave dynamics. 
The formalism allows one to efficiently treat 
radiation from complex sources, often having a significant nondeterministic, 
statistical character. 

The method introduced below exploits a connection between the 
field-field \emph{correlation function} (CF) and the WDF \cite{Littlejohn_1993,Dittrich_2009}.
Both quantities have been studied intensively in the 
physics and optics literature. For wave chaotic systems, Berry's conjecture postulates
a universal CF equivalent to correlations in Gaussian random fields
\cite{Berry01JPA_1977,Hemmady2012}. Non-universal corrections can be
retrieved by linking the CF to the Green function of the system
\cite{Creagh_1997,Hortikar_1998,Weaver_2001,Urbina_2006}. 
In this paper, we describe how field-field CFs can be efficiently 
propagated using ideas based on ray propagation in phase-space. We discuss 
furthermore non-paraxial effects as well as including near field effects due to 
evanescent wave contributions. A systematic expansion of the Wigner function 
propagator including next-to-leading-order effects in the propagating regime 
leads to Airy-function integral kernels containing the ray-tracing propagator in 
the small wavelength limit, akin to the treatments in \cite{Marcuvitz_1991, Littlejohn_1993}.  
We also show that our WDF representation confirms 
the validity of the generalised form of the Van Cittert-Zernike (VCZT) theorem discussed in 
\cite{Cerbino_2007}. We give a natural extension of this generalised VCZT  for non-paraxial sources 
and in the near-field region where evanescent waves play a prominent role.

We illustrate these techniques in the context of  
applications in EMC and related issues. 
Here, the system under investigation represents
a high-density interconnect of integrated electronic circuits. Simulating
EM field distributions in a reliable way is highly topical; in addition, the 
wave correlation function can be measured explicitly in this regime 
\cite{Russer_2012}, thus providing the necessary input information for 
numerical simulations. The system under consideration
in this paper consists of a series of parallel tracks carrying partially correlated currents, 
and mimics the typically very complex EM sources found on
PCBs. The method has much
wider application, however. In particular, when combined with 
fast phase-space propagation methods such 
as the {\em Discrete Flow Mapping} techniques 
developed in the context of vibro-acoustics \cite{Tanner2014547, Chappell2013}, the 
proposed WDF approach offers an ideal platform for developing a 
universal high-frequency simulation method. 

\section{Phase-space representation of classical fields}
Radiation from simple EM sources such as antennae can be characterised  
deterministically through classical electrodynamics methods \cite{jackson1998}. 
Even though such sources are regular and homogeneous, efficiently predicting
far-field emission from the near-field pattern requires 
non-trivial effort if the sources are extended over many wavelengths \cite{Bucci_1987}. 
EM sources are becoming increasingly complex, however, and the problem
of radiation from digital circuits or PCBs presents
even greater challenges.
Modelling such sources deterministically is often infeasible due
 to the complexity of the structures, whose details may not even be known in practice.
Each component of such a complex EM source is typically driven 
by unknown sets of quasi-random voltages, subject to fast transients 
\cite{Arnaut2013}. 
This is due to the presence of a multitude of electronic components whose
switching behaviour {depends} on the instantaneous operation mode of 
the circuit, and whose excitation 
signals are intrinsically random, or highly sensitive to 
frequency \cite{Russer_2012}. 
Consequently, the physical investigation of these scenarios challenges 
existing analytical and numerical techniques, 
and calls for more sophisticated modelling tools. 

It is thus natural to use statistics as a language for describing 
the radiation from such complex sources. Specifically, we do not
attempt to characterise or propagate the field itself, which is
typically hard to obtain in practice, but
rather its two-point CF. It has been demonstrated in
\cite{Russer_2012} that corresponding measurements are feasible 
in the context of emission from electronic devices and PCBs.
Here we describe the basic elements needed to use such
measurements as input for a practical algorithm with which to predict field intensities
and correlations away from the source. Initially we consider
radiation into free space in Sec.\ \ref{FreeSpaceSec} by studying a simple 
model source and a more realistic source obtained from a full field simulation. 
In Sec.~\ref{ReflectionSection}, we describe an
application to a problem with reflecting boundaries, which is a first step
towards our ultimate goal of extending the method to propagation of 
CFs in more complex environments such as cavities and larger structures.

We start from a planar source at $z=0$, parametrized by
coordinates $\x=(x_1,\cdots x_d)$ with $d=1$ or 2 in general, 
and radiating into  in the half-space $z>0$. 
We  aim to predict the CF
\begin{equation}\label{eqn:corr}
\Gamma_z (\x_B,\x_A) = 
\langle\psi(\xx_B,z)\psi^*(\xx_A,z) \rangle
\end{equation}
for $z>0$ under the assumption than it can be measured 
(or otherwise modelled) near the source screen $z=0$. Here, $\langle.\rangle$ denotes
an ensemble average over different source field correlations such as a time or 
frequency-band average.
Furthermore, $\psi(\x,z)$ denotes one of the tangential field components 
in the frequency domain. The results easily extend to
cross-correlation between different components.

In the past, the focus has often been on predicting the propagation 
of probability density functions (PDF) of waves 
passing through time-domain random \cite{Shechao_1988} or turbulent 
\cite{Crisanti_1993} media. 
In our approach, the propagation itself is treated deterministically,
whereas the radiation from the source is characterised statistically. This can be
done,
for example, by measuring the spatial field along a surface close to the
source and determining the source CF by 
averaging the signal over time. 
We thereby eliminate statistical fluctuations carried by the wave fields by 
ensemble averaging physical observables over suitable parameters. 

We now present the CF propagation rule explicitly for single field components. 
Polarisation effects can also be accounted for by propagating the 
field-field correlation tensor, which can be derived from the dyadic free-space 
Green's function \cite{Alonso_2004, Luis_PRA_2007}.

The field in the region $z>0$
is naturally presented in terms of
the partial Fourier transform,
\[
{\phi} \left ( \p, z\right ) = \int \e^{-\rmi
  k\pp\cdot\xx} \psi(\xx,z)\d\x,
\]
where $\psi(\x,z)$ denotes a field
component on the screen itself ($z=0$) or to its right ($z>0$) and $k$ is the wave vector.
The radiated fields can then be reconstructed using the evolution of this partial 
field. This can be calculated by using the dyadic second Green identity which, in 
a source-free region, becomes the dyadic version of  Huygen's principle \cite{harrington_book}. 
Being a convolution integral, the partial Fourier transform of the
surface integral {transforms to} an algebraic equation.
Then, the boundary conditions given by the fields sampled in the near-field region of the source
can be used to eliminate the magnetic field in {such an equation}. The result of this procedure, 
restricted to the {electric field components} parallel to the source plane, is 
the following inhomogeneous plane-wave solution
\begin{equation}\label{eqn:phi_p_z}
 {\phi} \left ( \p, z\right ) = \e^{\rmi k z {T}(\p)}{\phi} \left (\p, 0 \right ) \, ,
\end{equation}
where 
\begin{equation}\label{eqn:Tp}
 {T} \left ( \p \right ) = 
\left\{
\begin{array}{ll}
\sqrt{1 - |p|^2} & \mbox{if $|p|^2<1$}\\
\rmi\sqrt{|p|^2-1} & \mbox{if $|p|^2>1$}.
\end{array}
\right.
\end{equation}
For the moment, we neglect waves incident from the right and thus only describe
radiation from a strong, directional source; including incoming waves at the interface
can be introduced formally using the boundary integral equations
according to the discussion in \cite{Creagh2013}. 
An example of this scenario, 
involving a planar reflector beyond the source,  
will be given in Sec.~\ref{ReflectionSection}.
Here, $\p=(p_1, \ldots, p_d)$  takes the meaning of a momentum tangential to 
the $d$ dimensional source plane. In the ray-dynamical limit, we may identify
\begin{align}
\left | \p \right | & =  \sin\alpha \,\, , \label{eqn:momentum} \\
T \left ( \p \right ) 
 & \equiv p_z 
= \cos \alpha \,\, , \label{eqn:momentumz}
\end{align}
where the angle $\alpha$ describes the direction of the ray with respect to the local outward normal 
to the source. 
In this perspective, ${T} \left ( \p \right )$ represents
a generalized kinetic energy of the ray. The case $|p|^2>1$ in
(\ref{eqn:Tp}) corresponds to evanescent propagation, which does not
contribute to the far-field, but may be detectable in the near field; see also the discussion
in Secs.\ \ref{whitenoise}, \ref{tracks}. In order to represent wave fields in phase-space using canonical
coordinates $(\x,\p)$ parallel to the  source plane, we define the WDF
\begin{equation}\label{eqn:Wigner_Weyl_transf}
 \begin{split}
  W_z \left ( \x, \p \right ) =
\int\e^{-\rmi k\p\cdot\s}\Gamma_z(\x+\s/2,\x-\s/2)\d \s = \\ 
  \left(\frac{k}{2 \pi }\right)^d
  \int \, \e^{\rmi k \x \cdot \q} \, 
  \left\langle {\phi} \left ( \p + \q/2, z \right ) {\phi}^{*} \left ( \p - 
  \q/2, z \right ) \right\rangle \, \d \q .
 \end{split}
\end{equation}  
Upon insertion of (\ref{eqn:phi_p_z}) in
(\ref{eqn:Wigner_Weyl_transf}), and by exploiting the 
inverse transformation to represent the source correlation (at $z=0$)
in terms of the source Wigner function $W_{0} \left ( \x, \p \right )$, we find
\begin{equation}\label{eqn:WDF_prop}
  W_z \left ( \x, \p \right ) = 
  \int \, \mathcal{G}_z \left ( \x, \p, \x', \p' \right ) \, W_0 \left ( \x', \p' \right ) \, 
  \d\x'\d\p' . 
\end{equation}
This provides us with a propagator of the Wigner function taking the form 
\begin{eqnarray}\label{eqn:Green_int_op}
  &&\mathcal{G}_z \left ( \x, \p, \x', \p' \right ) = \left(\frac{k}{2 \pi}\right)^d \delta \left ( \p - \p' \right ) \\
 && \times \int \, \e^{\rmi k \left ( \x - \x' \right ) \cdot \q + \rmi k z 
   \left ( {T}(\p + {\q}/{2}) - {T}^*(\p - {\q}/{2}) \right )} \, \d\q ,\nonumber
\end{eqnarray}
where the $\delta$-function represents translational invariance in $\x$ and the
corresponding conservation of momentum. 
Eq.\ (\ref{eqn:Green_int_op}) 
provides a scheme to propagate wave densities in phase-space 
for arbitrary sources, no matter how complex or rapidly varying. 
The propagation of the correlation functions themselves
can subsequently be retrieved by an 
inverse Fourier transform of (\ref{eqn:WDF_prop}). That is,
\begin{equation}\label{eqn:Corr_from_WDF}
  \Gamma_z(\x_B,\x_A) = \left ( \frac{k}{2 \pi} \right )^{d} \int\e^{\rmi k \s\cdot\p}\, W_z \left ( \x, \p \right ) \d \p,
\end{equation}
where $x = (x_A + x_B)/ 2$, and $s = x_B - x_A$.
The intensity $I_z$ as function of the distance $z$ can be retrieved using \cite{torre2005}
\begin{equation}\label{eqn:pos_int}
 I_z(x) = \Gamma_z(\x,\x) = \left ( \frac{k}{2 \pi} \right )^{d} \int W_z \left ( \x, \p \right ) \d \p.
\end{equation}

\section{Ray Tracing Approximations}
Asymptotic approximation of the propagator
(\ref{eqn:Green_int_op}) leads to a direct propagation method 
for the WDF in terms of rays \cite{Berry_Wigner_1977,
Littlejohn_1993,Rios_2002,Dittrich_2006, Alonso_2011}. 
We will give a derivation of this ray limit below and will also discuss 
more subtle wave effects such as evanescent 
decay into the near-field and higher order (in $1/k$) wave corrections.

The simplest ray-based approximation is obtained under the
assumption that the CF is quasihomogeneous at the source,
that is, $\Gamma_0(\x_B,\x_A)=\Gamma_0(\x+\s/2,\x-\s/2)$ 
varies with $\x$ on a
larger-than-wavelength scale. In that case, significant contributions to
(\ref{eqn:Green_int_op}) are obtained only for small $\q$ and we
can expand the phase difference $\Delta{T} \left (\p, \q \right ) 
= {T}(\p + {\q}/{2}) - {T}^*(\p - {\q}/{2})$ around $\q = 0$. 

In the region $|p|^2<1$ corresponding to propagating waves, 
the difference $\Delta T$ receives contributions only from odd powers 
of $\q$. Neglecting cubic and higher order terms we find that
\begin{equation}\label{eqn:Frobenius_Green}
\small
 \mathcal{G}_z \left ( \x, \x'; \p, \p' \right ) 
\approx 
 \delta \left ( \x - \x' - \frac{z\p}{T(\p)}
 \right ) \,\delta \left ( \p - \p' \right ) \, .
\end{equation}
This is the Frobenius-Perron (FP) propagator \cite{Ott_2002} for radiation into free space
and leads to the evolution \cite{Littlejohn_1993}
\begin{equation}\label{eqn:Frobenius_WDF}
 W_z \left ( \x, \p \right ) \approx 
W_0 \left ( \x - \frac{z\p}{T(\p)}, 
\p \right )
\end{equation}
of the WDF in the region $|p|^2<1$. 
This approximation is equivalent to identifying the propagation of the
WDF with the propagation of phase space densities along rays according
to  the evolution
\begin{eqnarray}\label{eqn:ray_dyn}
 \x &=& \x' + \frac{z\p}{T(\p) }
\\
 \p &=& \p'\, .\nonumber
\end{eqnarray}
The paraxial approximation is obtained
by using the linearized flow in the regime $|p|^2\ll 1$ \cite{Alonso_2011}. Outside this
regime, the full flow has asymptotes along $|p|^2=1$, the transition to evanescent propagation.
Note that inserting Eq.\ (\ref{eqn:Frobenius_WDF}) in Eq.\
(\ref{eqn:Corr_from_WDF})
and including only the contributions from the propagating region
$|p|^2 <1$  leads to  
\begin{equation} \label{soso} 
 \Gamma_z \left ( \x_B, \x_A \right )\approx
\left ( \frac{k}{2 \pi} \right )^d
\int  \e^{\rmi kp s} W_0
  \left ( x - \frac{zp }{T(p)}, p \right ) \d p 
\end{equation} 
In the paraxial regime, that is for $|p|^2 \ll 1$, and for quasi-homogeneous sources, Eq. (\ref{soso}) 
retrieves the well-known van Cittert-Zernike theorem (VCZT) or generalisation thereof
\cite{Winston_2002, Cerbino_2007}. In the next section we will show how to extend the VCZT to the 
non-paraxial regime and including  evanescent waves.

Expanding $\Delta T$ further in $\q$ leads to an improved propagator
which is capable of mapping less homogeneous CF's. Including
cubic terms leads in the 2D case, for example, to the Airy form
\begin{equation} \label{airy_approx}
 \mathcal{G}_z \left ( \x,\p,\x',\p'\right ) \approx \delta_a\left(\x-\x'-\frac{z\p}{T(\p)}\right)\delta(\p-\p'),
\end{equation}
where
\[
\delta_a(u) = a\textrm{Ai}(au),
\]
with  $a=2k/(kzT'''(p))^{1/3}$ and Ai denotes the Airy function. Note
that $\lim_{|a|\to\infty}\delta_a(u) = \delta(u)$, so the
FP form is obtained in the limit of large wavenumber $k$ as 
expected. Similar results have been obtained in the context of 
the propagation of EM waves through 
inhomogeneous media \cite{Marcuvitz_1991}.  

Further improvements over the basic FP propagation
(\ref{eqn:Frobenius_Green}) are obtained by accounting for evanescent
decay into the near-field, which emerges from contributions 
$|p|^2>1$ in Eq.\ (\ref{eqn:WDF_prop}). 
Since the kinetic operators in Eq.\ (\ref{eqn:Green_int_op}) now add
constructively,  the leading contribution is formed by the zeroth 
order term in the expansion of $\Delta T(\p,\q)$ and we obtain
\begin{equation}\label{eqn:WDF_evan}
  W_z \left ( \x, \p \right ) \approx \e^{- 2 k z \sqrt{|p|^2 - 1}} 
\, W_0 \left ( \x, \p \right ), \qquad |p|^2 > 1 .
\end{equation}
Improved approximations may be achieved by treating
the exponent beyond leading order, but we find that (\ref{eqn:WDF_evan}) gives a good
description of evanescent decay already as discussed in the next section.

\section{Radiation into Free Space}\label{FreeSpaceSec}

We now test the effectiveness of the FP propagator in the simple case of
radiation into free space. 
The first example treated in Sec.~\ref{Gauss-Schell} assumes a quasi-homogenuous  source
distributed according to the Gauss-Schell model. We will examine the near-field 
behaviour in more detail in Sec.\ \ref{whitenoise}, considering in particular 
the limit of completely uncorrelated sources. In a second example
in Sec.~\ref{tracks}, we consider a more complex set-up mimicking the
more realistic sources expected in typical EM applications.
We will restrict ourselves in these examples to 2D models (so $d=1$) 
and characterise the behaviour of field-field correlations by focusing on the propagation 
of {\em one} field component along $z$. We select the field tangent to the source. 

\subsection{Propagation of Gauss-Schell model} \label{Gauss-Schell}

\begin{figure}
 \centering
 \includegraphics[width=7.5cm]{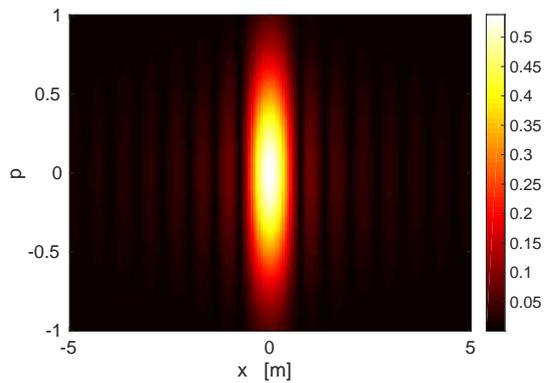}
 \caption{\label{fig:fig1w} Magnitude of the WDF of  a 1D Gauss-Schell correlation function. The radiation frequency is 
 $f=1$ GHz corresponding to $\lambda = 0.3$ m. The distribution widths are $\sigma_x = 1.0$ m, and $\sigma_s = 0.1$ m.}
\end{figure}

Using a simple 2D model for the emission of partially coherent EM radiation, we assume a source correlation in terms of
a truncated 1D Gaussian-Schell model 
\begin{equation}\label{eqn:Corr_trunc_Gaus}
 \begin{split}
  \Gamma_0 \left ( x_B, x_A \right ) = I_0 \exp \left [ - \frac{\left ( x_B - x_A \right )^2}{2 \sigma^2_s} \right ] \, \exp \left [ - \frac{\left ( x_B + x_A \right )^2}{8 \sigma^2_x} \right ] \, \\ 
  \times \chi_l \left ( x_B \right ) \, \chi_l \left ( x_A \right ) \, .
 \end{split}
\end{equation}
where  $l$ is the length of the source.
Here, the characteristic functions 
\begin{equation}
 \chi_l \left ( x \right ) = 
 \begin{cases}
 1,&\qquad \left |x \right | \leq \frac{l}{2},\\
 0,&\qquad \left |x \right | \geq \frac{l}{2}
 \end{cases}
\end{equation}
account for the finite size of the source. 
The quasi-homogeneity condition can be expressed through demanding 
$\sigma_s \sim \lambda\ll \sigma_x$, {where $\lambda= 2 \pi/k$} is the optical wavelength. 
The source WDF is then found to be 
\begin{equation}\label{eqn:WDF_trunc_Gaus}
 \begin{split}
  W_0 \left ( x, p \right ) = 
     I_0 \exp \left [ - \frac{x^2}{2 \sigma^2_x} \right ] \sqrt{\frac{\pi}{2}} \sigma_s \exp \left ( - \frac{k^2 p^2 \sigma^2_s}{2} \right ) \\ 
     \left [ {\rm erf} \left ( \frac{l - 2 \left | x \right |}{\sigma_s \sqrt{2}}
    - \rmi \frac{k p \sigma_s}{\sqrt{2}} \right ) - {\rm erf} \left ( - \frac{l - 2 \left | x \right |}{\sigma_s \sqrt{2}} - \rmi \frac{k p \sigma_s}{\sqrt{2}} \right ) \right ] \, .
  \end{split}
\end{equation}
For extended sources, for which $l \gg \lambda$, and for $\x$ inside
the region occupied by the source, Eq.\ (\ref{eqn:WDF_trunc_Gaus})
simplifies to
\begin{equation}\label{eqn:WDF_Gaus}
 \begin{split}
  W_0 \left ( x, p \right ) \approx 
     \sqrt{2\pi} \sigma_s I_0 \exp \left [ - \frac{x^2}{2 \sigma^2_x} - \frac{k^2 p^2 \sigma^2_s}{2} \right ]. 
  \end{split}
\end{equation}
Fig.\ \ref{fig:fig1w} shows the WDF in phase-space at $z=0$ for a spatially extended source \cite{Bastiaans1979}.
Here, and in all other computations with the Gauss-Schell model, we work at a frequency of operation of $1$ GHz corresponding to  $\lambda = 0.3$ m and choose
$\sigma_x = 1.0$ m, $\sigma_s = 0.1$ m. 

\begin{figure}
 \centering
 \includegraphics[width=9.2cm]{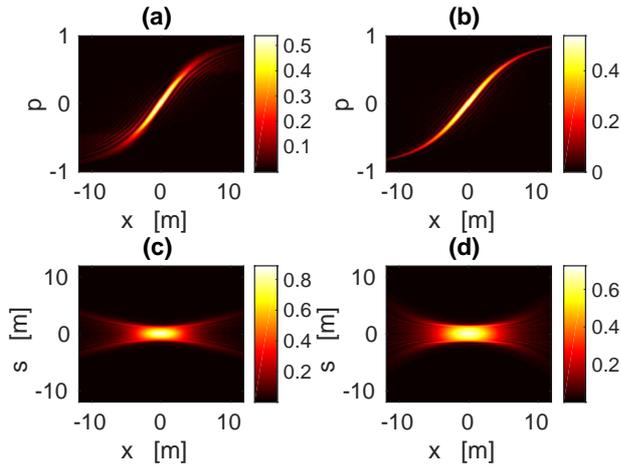}
 \caption{\label{fig:schell-exact} Propagated 
 Wigner distribution function (a), (b) and correlation function (c), (d) of a partially-coherent (1D Gauss-Schell) near-homogeneous source: 
 {exact ((a) and (c)) vs approximate - Frobenius-Perron - ((b) and (d))} computation at $z=10 \lambda$. The radiation frequency is 
 $f=1$ GHz. The distribution widths are $\sigma_x = 1$ m, and $\sigma_s = 0.1$ m.}
\end{figure} 
Fig.\ \ref{fig:schell-exact} shows the propagation of (\ref{eqn:WDF_trunc_Gaus}) as computed through the full integral operator (\ref{eqn:WDF_prop}) 
together with the propagation obtained by the Frobenius-Perron approximation (\ref{eqn:Frobenius_WDF}).
There is surprisingly good agreement between the exact and approximate behaviour even far from the paraxial regime. 
This is remarkable given that the ray tracing approximation is only valid to leading order. This constitutes a major computational 
advantage as the FP approximation reduces an integral equation to a simple coordinate transformation.  
The overall behaviour shown in Fig.\ \ref{fig:schell-exact}(a) and Fig.\ \ref{fig:schell-exact}(b) reflects the distribution shearing due to the geometrical 
ray propagation based on Eq.\ (\ref{eqn:ray_dyn}); see also \cite{Alonso_2011}. The CFs can now be obtained  by a back transformation according to Eq.\ (\ref{eqn:Corr_from_WDF}) and are shown in Fig.\ \ref{fig:schell-exact}(c) and Fig.\ \ref{fig:schell-exact}(d).

\subsection{Non-paraxial Van Cittert-Zernike theorem}\label{whitenoise}

In the following, we will focus on near-field effects for small distances from the source as a function of the source
correlation parameter $\sigma_s$. We are in particular interested in how the correlation length propagates in the near- 
field before reaching the linear VCZT regime. 

In the near-field limit, the WDF shows exponentially decaying 
evanescent components according to (\ref{eqn:WDF_evan}), while the WDF
remains essentially unchanged for the propagating part $|p|^2<1$. 
This leads to a model for the WDF with source distribution (\ref{eqn:WDF_Gaus}) of the form
\begin{equation}\label{eqn:Wz_inc_nf}
 \begin{split}
  W_z(x,p) \approx 
     \sqrt{2\pi} \sigma_s I_0 \exp \left [ - \frac{x^2}{2 \sigma^2_x}
       - \frac{k^2 p^2 \sigma^2_s}{2} \right ]\\ 
  \times\left\{
  \begin{array}{ll}
  1 & \mbox{if $|p|^2<1$}\\
  \e^{- 2 k z \sqrt{|p|^2 - 1}} 
   & \mbox{if $|p|^2>1$}\, .
  \end{array}
  \right.
 \end{split}
\end{equation}
Far enough from the source, such that 
evanescent components have completely decayed, while close enough that
evolution in the propagating region of phase space can still be
neglected, we model the WDF using

{
\begin{equation}\label{eqn:Wz_inc_ff}
 \begin{split}
  W_z(x,p) \approx 
     \sqrt{2\pi} \sigma_s I_0 \exp \left [ - \frac{x^2}{2 \sigma^2_x}
       - \frac{k^2 p^2 \sigma^2_s}{2} \right ]\\ 
  \times
  \left\{
  \begin{array}{ll}
   1 & \mbox{if $|p|^2<1$}\\
   0 
   & \mbox{if $|p|^2>1$}.
   \end{array}
  \right.
 \end{split}
\end{equation}
Using the inverse Fourier transform, Eq.\  (\ref{eqn:Corr_from_WDF}), we now obtain the correlation function from 
the WDF given by Eq.\ (\ref{eqn:Wz_inc_nf})
or Eq.\ (\ref{eqn:Wz_inc_ff}) acccordingly.
In Fig.~\ref{fig:corr_fnct} we show the resulting near-field evolution
of the correlation
function, placing the midpoint $\x=(\x_A+\x_B)/2=0$ at the centre of
the source.

\begin{figure}
        \centering
        \includegraphics[width=7.5cm]{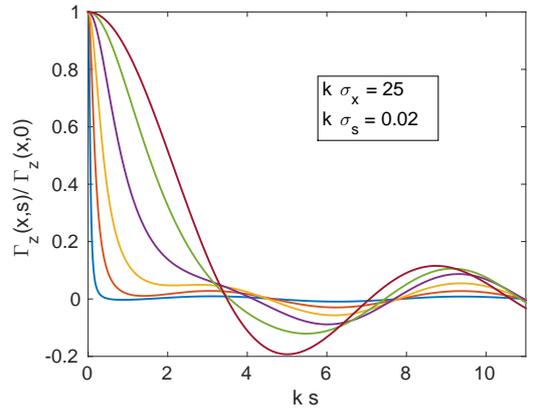}     
        \caption{\label{fig:corr_fnct} Evolution of the correlation function along the line  $x=0$ for a selected (partially) correlated source, 
        from $k z = 0.42$ (blue solid line) to $k z = 1.68$ (brown solid line).
        In the presence of evanescent waves the correlation width increases rapidly before the correlation function becomes 
        a sinc function, whose width then increases linearly according to the VCZT.
        }
\end{figure}

One observes that in the near-field regime, the width $\Delta s$ of
the correlation function  is smaller than $\lambda$, but it increases
rapidly towards $\lambda$ as  $z$ approaches and exceeds $\lambda$. 
The second moment of the correlation function is not defined and
cannot therefore be used to define a correlation length. Instead,
we define the correlation lengths to be the spacing at which the
correlation has fallen by a factor $1/\sqrt{\e}$: 
\begin{equation}\label{eqn:th_gaus}
  \Gamma_z (x+\Delta s/2,x+\Delta s/2)/ \Gamma_z (x, x) = \e^{-1/2}. 
\end{equation}
Note that for a Gaussian correlation function such as assumed for the
source in (\ref{eqn:Corr_trunc_Gaus}), this definition coincides
with the standard variance: $\Delta s = \sigma_s$.

With the definition adopted above, we can now obtain the 
correlation lengths from exact  wave propagation calculations. The results are 
shown in Fig.\ \ref{fig:Ds_VCZT} as a function of 
the distance $z$ for different source correlation lengths $\sigma_s$. The universal 
regime is shown as the blue dashed line. The VCZT regime starts around $kz > 1$.
\begin{figure}
       \centering
      \includegraphics[width=7.5cm]{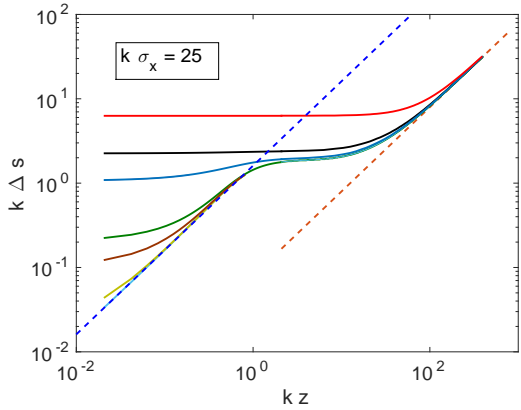}      
       \caption{\label{fig:Ds_VCZT} {Evolution of the correlation length $\Delta s$
       with z on a log-log scale at $x=0$ and for a {partially  
        correlated} 1D source with $k \sigma_s=0.002$ (cyan solid line)$ \ldots 6.3$ (red solid line). 
        Evanescent waves drive the rapid spatial (blue dashed line) increase preceding the 
        VCZT regime (red dashed line). A plateau is observed between $kz \approx 1$ and the 
        \emph{Rayleigh} range discovered in \cite{Cerbino_2007,Gatti_2008}.}
        }
\end{figure}

From Eqs. (\ref{eqn:Wz_inc_nf}) and (\ref{eqn:Wz_inc_ff}), one can
estimate the growth 
rate both in the near and the far field. Including non-paraxial effects,  there are three different regimes: 
\begin{itemize}
\item[i)] in the deep near field, with $kz< 1$ and $k \sigma_s < 1$, 
the correlation length increases linearly with a slope that is
independent of the frequency as well as of $\sigma_s$ and $\sigma_x$; 
\item[ii)] a no-growth regime with $\Delta s = const$. 
\begin{itemize}
\item For $k\sigma_s < 1$, one finds $k \Delta s \approx 1$  in the range 
$1 < k z< k \sigma_x$;\\
\item For $k \sigma_s > 1 $, then $\Delta s = \sigma_s$  in the range
$0<z< k \sigma_x$. 
\end{itemize}
The latter regime has already been described by Cerbino for 
paraxial sources \cite{Cerbino_2007}; 
\item[iii)] the VCZT regime for large $z$ with a linear growth of the correlation length according to 
\begin{equation}
\Delta s  \propto \frac{\lambda}{\ell} z \,,
\end{equation}
with a slope depending on the ratio of  wavelength to source dimension $\ell$.
\end{itemize}

We now motivate these three regimes in more detail, beginning with
case (i), which corresponds to $k\sigma_s< 1$, $kz < 1$. The
WDF $W_z(\x,\p)$ described by (\ref{eqn:Wz_inc_nf}) then decays slowly
along the $p$ axis as
$|p|$ increases beyond the propagating region $|p|^2<1$. In the extreme
nearfield the correlation function is proportional to the inverse
Fourier transform
of the function
\[
W_z(x,p) \sim \e^{-2kz |p|}
\]
of $p$, that is, 
\[
\Gamma_z(s) \sim \frac{2kz/\pi}{(2kz)^2+(ks)^2}.
\]
The correlation length defined by Eq.\  (\ref{eqn:th_gaus}) then takes
the form
\begin{equation}\label{proptoz2} 
 \Delta s \approx 2 \sqrt{\sqrt{\e}-1} \; z \approx  1.6109\,  z .
\end{equation} 
That is, we find in regime (i) that evanescent decay of the
sub-wavelength correlations in the source dominates in such a way that 
there is a \emph{universal} growth rate in the correlation 
length. The numerical value of the slope in (\ref{proptoz2})
is particular to the form taken in Eq.\  (\ref{eqn:th_gaus})
for the correlation length, but the qualitative conclusion 
applies more generally. The presence of evanescent
waves thus leads to a rapid increase of the correlation length in the near
field in this regime.  
This is  important for sources that show fluctuations on scales smaller than 
the wavelength, such as in the case of a fully uncorrelated 
{source $\sigma_s =0$, which} 
may serve as a model for  thermal sources \cite{Carminati_1999}. 

The plateau behaviour corresponding to regime (ii) arises when
$z$ is sufficiently large that (\ref{eqn:Wz_inc_ff}) describes 
the WDF, while $k\sigma_s\ll 1$. The correlation function 
is then proportional to the inverse
Fourier transform
\[
\Gamma_z(s) \sim \frac{1}{\pi} {\rm sinc} (ks)
\]
of the function
\begin{equation}\label{step}
W_z(x,p) \sim 
 \begin{cases}
 1,&\qquad |p|^2<1,\\
 0,&\qquad |p|^2>1
 \end{cases}
\end{equation}
of $p$. In this case the correlation length defined by Eq.\  (\ref{eqn:th_gaus}) takes
the form
\begin{equation}\label{proptoz}
 k \Delta s \approx 1.6443 
\end{equation} 
independent of $\sigma_s$.
It should be noted that if the condition  $k\sigma_s <1$ is
breached, then the Gaussian decay in $p$ present in
(\ref{eqn:Wz_inc_ff}) becomes the dominant feature and instead a
limiting plateau level
\[
\Delta s = \sigma_s 
\]
occurs, see Fig.\ \ref{fig:Ds_VCZT}. Note that in this case the plateau extends all the way to $z=0$ and the
linear regime of case (i) is not seen.

Finally, regime (iii) applies once evolution of the phase space takes
effect in the propagating region $|p|^2<1$. Assuming the
quasihomogeneous case $\sigma_x\gg l$ and considering first
only the near-field region $k\sigma_s\ll kz \ll 1$, 
then for a given midpoint $x$ the 
finite size of the source reduces the support in $p$ of the Wigner function and
(\ref{step}) is replaced by 
\[
W_z(x,p) \sim 
 \begin{cases}
 1,&\qquad \frac{x-l/2}{\sqrt{z^2+(x-l/2)^2}}<p<\frac{x+l/2}{\sqrt{z^2+(x+l/2)^2}},\\
 0,&\qquad \mbox{otherwise}.
 \end{cases}
\]
For simplicity consider the case $x=0$. Then the correlation function
obtained from the inverse Fourier transform of this function is
\[
\Gamma_z(s) \sim \frac{1}{\pi} {\rm sinc} \left(\frac{ks}{\sqrt{1+(2z/l)^2}}\right)
\]
and the correlation length defined by (\ref{eqn:th_gaus}) takes the form
\begin{equation}\label{proptozgen}
 \Delta s \approx 0.2617 \lambda\sqrt{1+(2z/l)^2},
\end{equation} 
generalising (\ref{proptoz}). In the farfield $z\gg l$, we find
\[
 \Delta s \approx 0.2617 \times \frac{2z\lambda}{l}
\]
(where the numerical prefactor is particular to the convention
(\ref{eqn:th_gaus})).
Alternatively, if $\sigma_x\gg l$, then the screen length becomes
unimportant and $\sigma_x$  provides the length scale appropriate to
the source intensity. An analogous calculation then allows us instead to
recover the basic form
\[
\Delta s \approx \frac{1}{2\pi} \times \frac{z\lambda}{\sigma_x}
\]
of the VCZT for $z\gg \sigma _x$.

\subsection{Application to a complex source}\label{tracks}
The field at the source at $z=0$ is often produced by 
a complex process such as tracks on a printed circuit board or integrated circuits in electronic devices; 
the radiation produced in the source region then propagates into free space. We model 
such a complex source here by a set of $N$ metallic wires driven by random time-domain voltages, as illustrated in Fig.\ \ref{fig:scheme}; 
a realization of the voltage $s_p \left ( t \right )$ driving a pin of the bundle is reported in 
Fig.\ \ref{fig:voltage} along with its spectrum $S_p \left ( f \right )$. 
\begin{figure}
        \centering
        \psfrag{pec}{\textbf{PEC}}
        \psfrag{x}{$x$}
        \psfrag{y}{$y$}
        \psfrag{z}{$z$}
        \psfrag{sp}{$s_p (t)$}
        \psfrag{d}{$d$}
        \psfrag{l}{$l$}
        \psfrag{hl}{$h_l$}
        \psfrag{hu}{$h_u$}
        \includegraphics[width=0.25\textwidth]{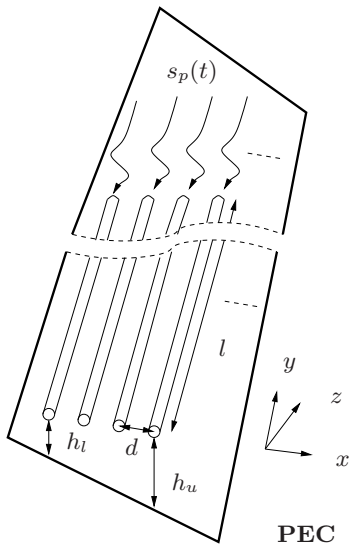}     
                \caption{\label{fig:scheme} Set of parallel metallic wires running above an oblique 
        perfect electric conductor (PEC) ground plane. 
        This complex source emits radiation in the half-space $z>0$. 
        The full-wave TLM simulation has been carried out for the configuration 
        $h_l=0.3025$ m, $h_u=0.358$ m, $l=1$ m, $d=0.06$ m.}
T\end{figure}
\begin{figure}[!t]
        \centering
        \includegraphics[width=0.49\textwidth]{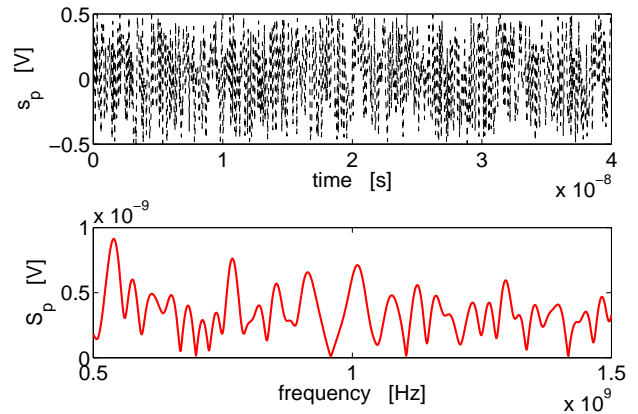}      
        \caption{\label{fig:voltage} Time- and frequency-domain behavior of the signal driving a pin of the bundle.}
\end{figure}
This represents a typical 
problem in EMC where one tries to obtain statistical information about 
an erratic signal. 

The presence of a perfect electric conductor along an oblique plane
makes the source radiate only into the half-space $z>0$: this
 mimics a configuration that is widely used in the design of printed circuit boards. 
We use $N=12$ wires very close to each other and to the metallic plane in terms of wavelength. 
Therefore, it is reasonable to think of this circuit as a collection of random sources of partially coherent radiation. 

The exact fields emitted from such a complex structure are computed through an in-house Transmission Line Matrix (TLM) 
code \cite{christopoulos1995transmission}. This is a time domain method for modelling 3D electromagnetic field interactions 
with complex structures that may include a variety of materials. The technique is based on the equivalence 
between electric and magnetic fields and the voltages and currents on a network of transmission lines. After discretizing space, 
the fields in individual cells are modelled by transmission lines incident from each cell-face and intersecting at the cell centre forming a junction. 
Each of these orthogonal transmission lines allows for the propagation of electromagnetic waves. The waves are characterised by voltage and 
current and their associated electric and magnetic fields. In order to obtain the desired correlation functions, we sample the numerically 
obtained fields in a plane above the tracks at different times in order to create a suitable ensemble of uncorrelated circuit realisations. 
This is used as a basis for calculating field-field correlation functions and their Wigner functions both in the near- and far-field. 

Figs.\ \ref{fig:FF_WDF} (a) and (b) show the comparison between the WDF as computed through the full-wave (TLM) 
simulations, and the WDF obtained by the FP approximation (\ref{eqn:Frobenius_WDF}) in the far-field 
at $z = 2.3 \lambda$. In the TLM calculation, the full time-dependent field is propagated out from the source, 
while in the FP approximation, the WDF obtained from the signal at the source (as shown in Fig.\ \ref{fig:voltage}, see 
also Fig.\ \ref{fig:NF_WDF} (a)) is propagated according to (\ref{eqn:Frobenius_WDF}).
There is good agreement between the behavior predicted from full-wave simulations and the FP approximation, even 
though the source exhibits strong inhomogeneities. 
Interestingly, Fig.\ \ref{fig:FF_WDF} shows the same Wigner distribution
shearing as in Fig.\ \ref{fig:schell-exact} following 
the geometrical interpretation (\ref{eqn:ray_dyn}) of the correlation propagation. 
It is worth stressing that such a Wigner function 
challenges the FP approximation
(\ref{eqn:Frobenius_WDF}), whose underlying assumption is
quasi-homogeneity. Note that we can always also switch to 
the exact transport rule (\ref{eqn:WDF_prop}), which is computationally
more expensive than the FP approximation, but still orders of magnitudes faster than a full TLM calculation.
Propagated CFs as shown in Fig.\ \ref{fig:FF_WDF} (lower plots (c) and (d))
are finally obtained by applying {the inverse Fourier transform (\ref{eqn:Corr_from_WDF}).}

Note that we also find
a pronounced broad side radiation around $p\approx \pm1$ (corresponding to $\alpha \approx \pm \pi / 2$), and 
a strong asymmetry of the Wigner distribution due to the oblique
metallic reflector. Those features can be captured by inspection of the WDF representation 
in phase-space, while they are less apparent in the propagated correlation function shown in
Fig.\ \ref{fig:FF_WDF} (c) and (d).

\begin{figure}[!t]
        \centering
        \includegraphics[width=0.49\textwidth]{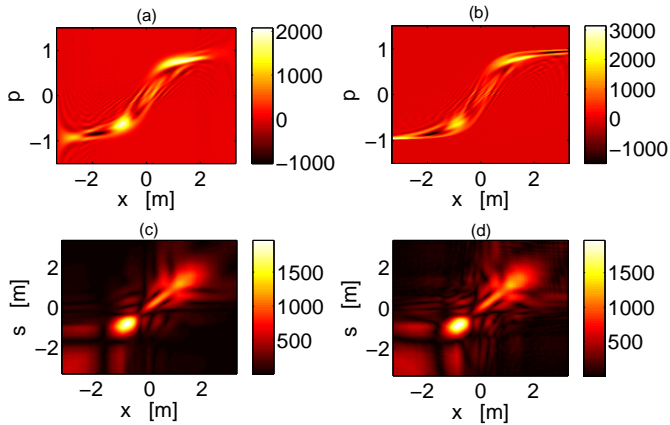}      
        \caption{\label{fig:FF_WDF} Far-field WDF (upper plots) and correlation functions (lower plots) at $z=2.3$ $\lambda$: 
        comparison between TLM computation (left column) and Frobenius-Perron analytical approximation (\ref{eqn:Frobenius_WDF}) 
        (right column). 
        The propagated correlation has been calculated through (\ref{eqn:Corr_from_WDF}).}
\end{figure}
\begin{figure}[!t]
        \centering
        \includegraphics[width=0.5\textwidth]{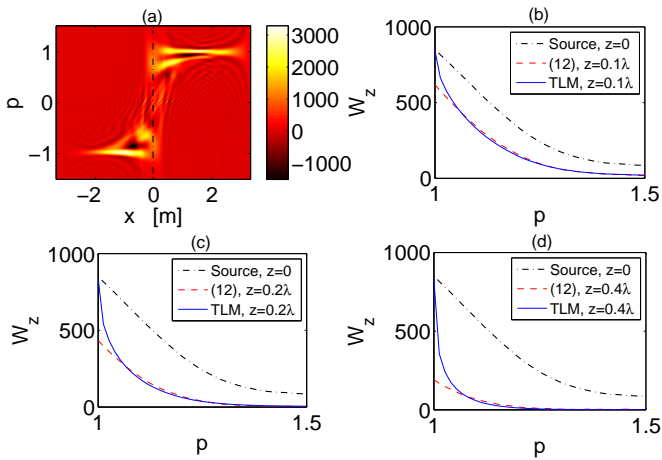}      
        \caption{\label{fig:NF_WDF} (a) Source distribution; near-field WDF: comparison between TLM computation and evanescent-wave approximation (\ref{eqn:WDF_evan}) for (b) $z=0.1 \lambda$; (c) $z=0.2 \lambda$ and (d) $z=0.4 \lambda$ along $x=0$.}
\end{figure}

The source distribution  $W_0(x,p)$ as obtained from the radiated signal, 
see Fig.\  \ref{fig:voltage}, is shown in Fig.\ \ref{fig:NF_WDF} (a). Note that 
the region with $|p|^2>1$ corresponds to evanescent contributions.
In Fig.  \ref{fig:NF_WDF} (b)-(d), a comparison between 
WDFs as computed through the full-wave (TLM) simulations and those obtained 
using the WDF propagator incorporating evanescent contributions, 
Eq.\ (\ref{eqn:WDF_evan}), are shown along the line $x=0$.
We find that propagation beyond $z=0.1\, \lambda$
results predominantly in an exponential reduction of the WDF in the 
region $|p|^2 > 1$. In the far-field, the radiation energy is restricted to the 
phase-space region $\left | \p \right | < 1$, as can be seen in the 
WDF in Fig. \ref{fig:FF_WDF} (a) and (b). 

A comparison of (TLM) simulated and (Frobenius-Perron) approximate 
far-field propagated energy
\begin{equation}
 E_z = \frac{1}{2} \, \epsilon_0 \, \left | \phi \left ( x \right ) \right |^2 = 
 \frac{1}{2} \, \epsilon_0 \, \int \, W_z \left ( x, p \right ) \, dp, 
\end{equation}
is shown in Fig.\ \ref{fig:en_den}. We see that the two numerical methods show 
qualitatively the same features, however, there are quantitative differences. We think that
these deviations are due to a difference in the numerical treatment of the boundary conditions at $x = \pm 3$ m. 
While the FP approach has no difficulties in treating these boundaries as completely open, 
the TLM method needs to model this with absorbing boundary conditions. These conditions
tend to be still slightly reflective, as is evident from the source distribution in Fig.\ 
\ref{fig:NF_WDF} (a) around $x = \pm 3$, $p= \pm 1$. This comparison highlights another 
advantage of the Wigner function propagation method. 
\begin{figure}[!t]
        \centering
        \includegraphics[width=7.5cm]{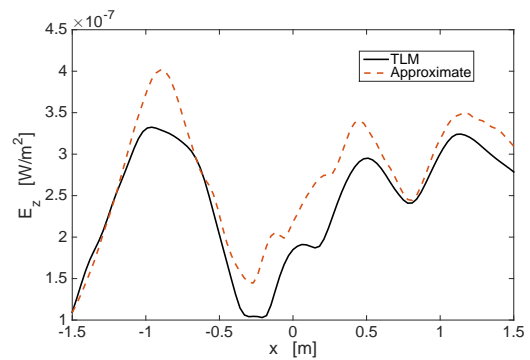}      
        \caption{\label{fig:en_den} TLM versus approximate energy density of the field radiated far from the source. 
               }
\end{figure}
The evaluation of the WDF of actual circuits can be done for the full electromagnetic field by using the approach 
described here component by component. 

\section{Reflection of partially correlated sources}\label{ReflectionSection}
{Having developed a framework for the propagation of CFs in free space, we are now interested in 
tackling the case of reflection from planar boundaries. 
In particular we would like to test the FP approximation in the presence of interference.}
It is then interesting to solve the canonical situation depicted in Fig.\ \ref{fig:Pla_ref}, where a planar reflector is 
located at distance $z=L$ from the source at $z=0$. 

The reflecting boundary is here for simplicity assumed to be 
parallel to the source plane, indefinitely extended in the $\hat{xy}$-plane, and made of an ideal 
perfect electric conductor (PEC). Therefore, for electric (TE) or magnetic (TM) fields perpendicular 
to $\hat{z}$, the Fresnel reflection coefficient reads $r \left ( \alpha \right ) = -1$, for all incoming 
angles $\alpha$ \cite{tsang2004}. We again consider for simplicity only a scalar field, or a 
single component of the vector field, emitted from the source. 
\subsection{Theory}
\begin{figure}
        \psfrag{x}{\small $\hat{x}$}
        \psfrag{y}{\small $\hat{y}$}
        \psfrag{z}{\small $\hat{z}$}
        \psfrag{s}{\small Planar source}
        \psfrag{d}{\small Scan plane}
        \psfrag{l}{\small Reflector}
        \psfrag{del}{\small $\Delta$}
        \psfrag{pec}{\small PEC}
        \psfrag{z0}{\small $z=0$}
        \psfrag{zd}{\small $z=D$}
        \psfrag{p1}{\small $\hat{\phi} \left ( \ushortw{p}_1, z \right )$}
        \psfrag{zl}{\small $z=L$}
        \psfrag{a1}{\small $\alpha_1$}
        \psfrag{d}{\small Detector}
        \centering
        \includegraphics[width=6.2cm]{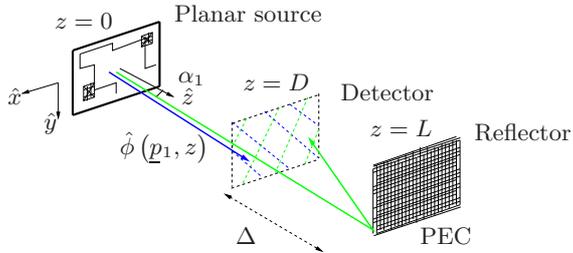}      
        \caption{\label{fig:Pla_ref} Arbitrary planar electromagnetic source emitting in the half-space $z>0$, in presence of a planar metallic boundary.}
\end{figure}

Consider a plane located at an arbitrary longitudinal coordinate $z=D$ between source and detector. 
The field distribution in the plane consists then of two contributions: the direct wave coming 
from the source, and the reflected wave bouncing off the reflector back to the source, that is, 
\begin{equation}\label{eqn:phi_p_dir_ref}
 \hat{\phi} \left ( p, z \right ) =  \e^{rmi k D T(p)}\phi \left (p, 0 \right ) 
 - \e^{\rmi k D T(p) + i 2 k \Delta T(p)}\phi \left (p, 0 \right ),
\end{equation}
where $\phi \left (p, 0 \right )$ is the field at the source plane $z=0$, $T(p)$ 
is defined as in (\ref{eqn:Tp}), and $\Delta = L-D$.  
The momentum space CF is formed as the product of the two fields in (\ref{eqn:phi_p_dir_ref}) 
and an ensemble average is taken as in Eq.\ (\ref{eqn:corr}). By plugging the closed-form expression (\ref{eqn:phi_p_dir_ref}) 
into the definition of the WDF (\ref{eqn:Wigner_Weyl_transf}), we find the phase space representation
\begin{equation}\label{eqn:Wigner_reflection}
\small
 \mathcal{W}_D \left ( p, x \right ) = W_D \left ( p, x \right ) 
 + W_{2L-D} \left ( p, x \right ) - \left [ W_{\Delta} \left ( p, x
   \right ) + \textrm{cc} \right ]  ,
\end{equation}
where the first two terms are direct and reflected contributions respectively, coming to the detector straight from the source or through 
the reflector, and the last two terms express the interference between direct and reflected waves with $\textrm{cc}$ 
standing for the complex conjugate. 

Following the procedure described in the previous subsection, it can be shown that direct and reflected terms in (\ref{eqn:Wigner_reflection}) 
can be calculated through the free-space propagation scheme in (\ref{eqn:WDF_prop}) and (\ref{eqn:Green_int_op}), 
with $z=D$ and $z=2L - D$ respectively, while the interference terms
lead to
\begin{equation}\label{eqn:WDF_prop_reflection}
 W_{\Delta} \left ( p, x \right ) = \iint \, \mathcal{G}_{\Delta} \left ( x, x'; p, p' \right ) \, 
 W_0 \left ( x', p' \right ) \, \d x' \d p' \,\, ,
\end{equation} 
with a modified Green integral operator 
\begin{eqnarray}\label{eqn:Green_int_op_reflection}
\small
  &&\mathcal{G}_{\Delta} \left ( x, x'; p, p' \right ) = \delta \left ( p - p' \right ) \frac{1}{(2\pi)^2} \\
 && \times \int \, \e^{\rmi k \left ( x - x' \right ) q + \rmi k D 
   \left ( T( p + \frac{q}{2}) - T^*( p - \frac{q}{2}) \right ) - 
   \rmi 2 k \Delta T^*( p - \frac{q}{2}) } \, \d q \nonumber \,\, .
\end{eqnarray}
For the class of statistically quasi-homogeneous sources, we may again expand the exponent in (\ref{eqn:Green_int_op_reflection}) 
in a Taylor series in $q$, and retain only terms up to first order.
This results in a Frobenius-Perron approximation of the interference terms, leading to a phase-factor of the optical length $\Delta$ besides the 
Dirac's delta in (\ref{eqn:Frobenius_Green}). 
Adopting the same linear approximation for each term in (\ref{eqn:Wigner_reflection}) gives the updated WDF 
\begin{eqnarray}\label{eqn:Frobenius_WDF_reflection}
& \mathcal{W}_D \left ( x, p \right ) \approx   
      W_0 \left ( x - \frac{Dp}{T(p)},p \right ) 
\\ \nonumber
  & + 
\left . 
W_0 \left(x - \frac{\left(2L - D \right )p}{T(p)}, p \right )\right. \\ 
  & 
- 2 \cos \left ( 2 k \Delta T \left ( p \right ) \right ) \, 
W_0 \left(x-\frac{Lp}{T(p)}, p \right ) \nonumber
.
\end{eqnarray}
Similar expressions have been found in quantum mechanics \cite{Littlejohn} and optics \cite{torre2005} 
for two overlapping wave-functions. 

Again, the propagated CF can be obtained by the inverse Fourier transform (\ref{eqn:Corr_from_WDF}) of (\ref{eqn:Wigner_reflection}) 
or (\ref{eqn:Frobenius_WDF_reflection}), the latter being closely related to the free-space VCZT. 
 
\subsection{Numerical results}
We chose again an initial correlation density distributed according to the Gauss-Schell model, Eq.\ (\ref{eqn:Corr_trunc_Gaus}), with 
corresponding source WDF shown in Fig.\ \ref{fig:fig1w}. We work as usual at a frequency of operation of $1$ GHz corresponding to  $\lambda = 0.3$ m and choose $\sigma_x = 1.0$ m, $\sigma_s = 0.1$ m. 

We further suppose a metallic mirror at $L=1.8$ m \textbf{($6 \lambda$)}.  
The propagation of the intensity from the source to the mirror can be found by evolving 
the source WDF with the exact rule composed of Eqs.\ (\ref{eqn:WDF_prop}), (\ref{eqn:Green_int_op}) and those for the 
interference terms,  Eqs.\ (\ref{eqn:Wigner_reflection})
-- (\ref{eqn:Green_int_op_reflection}), and then 
inverse Fourier transforming the propagated WDF according to Eq.\  (\ref{eqn:Corr_from_WDF}). 
The coherent energy $I_z(x)$ reaching the scan plane 
at $z=D$ is given by Eq.\ (\ref{eqn:pos_int}), that is, by considering the correlation function at $s = 0$.

Figure \ref{fig:fig2w} shows the behavior of the intensity $I_z (x=0)$ near the mirror, from $D=1.0$ m to $D=1.8$ m. 
The solid black line is computed through the full Green's integral operators (\ref{eqn:Wigner_reflection}) and (\ref{eqn:Green_int_op_reflection}), while 
the dashed red line is obtained by the Frobenius-Perron approximation (\ref{eqn:Frobenius_WDF_reflection}).
\begin{figure}
 \centering
 \includegraphics[width=7.5cm]{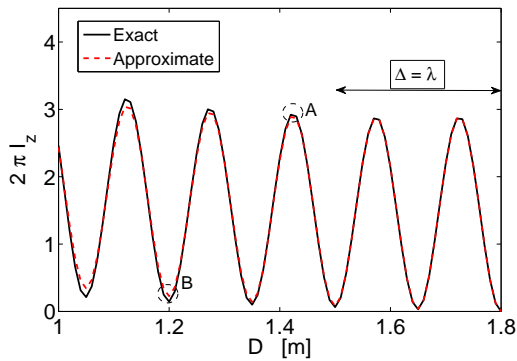}
 \caption{\label{fig:fig2w} Interference pattern formed by the intensity $I_z$ ($x=0$) along the longitudinal direction as the 
 scan plane approaches the reflector. Exact (black solid line) versus approximate (red dashed line) 
 computations are compared.}
\end{figure}
The oscillatory behaviour in (\ref{eqn:Frobenius_WDF_reflection}) is due to the interference terms in the WDF. 

In Figs.\ \ref{fig:refl_A} and \ref{fig:refl_B}, we show the magnitude of the WDF and the associated CFs at a distance $\Delta =
1.25 \lambda$ (position A in Fig.\ \ref{fig:fig2w}) 
and at a distance $\Delta = 2 \lambda$ 
(position B in Fig.\ \ref{fig:fig2w}) from the mirror, 
respectively. While good agreement between the exact and the approximate propagation using the FP approximation is 
achieved at position A, a maximum in the 
correlation function, the same is not true at position B.  Here the intensity is suppressed due to destructive interference and the magnitude
of the CF is itself only of order $O(1/k)$. To obtain the good agreement shown in Fig.\ \ref{fig:refl_B}, we need to take into 
account higher order corrections in the WDF propagator such as using the Airy function integral kernel, Eq.\ (\ref{airy_approx}). 
The improvement when going from the leading order FP to the Airy function approximation is shown in Fig.\ \ref{fig:refl_B} (b) 
to (c), which need to be compared with the exact WF Fig.\ \ref{fig:refl_B} (a); the corresponding propagated CF is displayed in 
Fig.\ \ref{fig:refl_B} (d). Only after 
going beyond the FP approximation in this way are we able to reconstruct the fine structure of the WDF. This finding is not 
surprising, but  remarkable nevertheless; computing WDFs in a multi-scattering environment will encounter exactly these problems and we 
have shown that the Airy-function approximation - still faster than a full WDF propagation - can handle  interference corrections 
successfully. We note that these corrections have been reported also in the ``diffusive'' Green function presented in \cite{Marcuvitz_1991}.

\begin{figure}[t!]
 \centering
 \includegraphics[width=0.51 \textwidth]{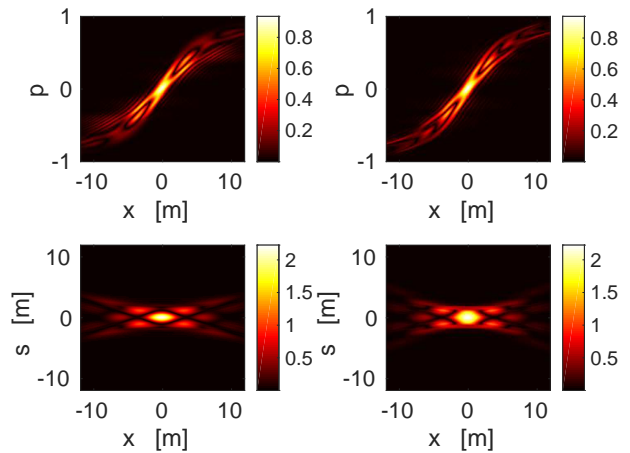}
 \caption{\label{fig:refl_A} Magnitude of the WDF of a 1D Gauss-Schell source: exact (left plots) versus approximate (right plots) computation at $\Delta=1.25 \lambda$ (position A in Fig.\ \ref{fig:fig2w}). Related correlation functions are reported in the lower plots.}
\end{figure}

\begin{figure}[t!]
 \centering
 \includegraphics[width=0.51 \textwidth]{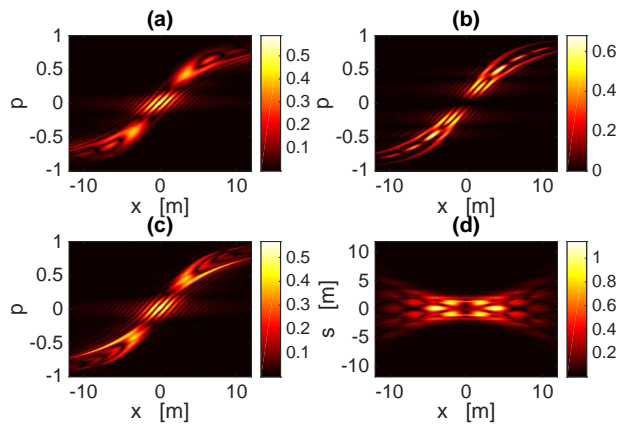}
 \caption{\label{fig:refl_B} Magnitude of the WDF of a 1D Gauss-Schell
   source: exact (a) versus approximate using only FP approximation
   (b) and using the Airy approximation (c)  at $\Delta=2 \lambda$
   (position B in Fig.\ \ref{fig:fig2w}). The correlation function corresponding to (c) is reported in (d).}
\end{figure}

\section{Conclusion}
An exact propagator has been derived for field-field correlation
functions of complex sources. It has been applied to a
problem mimicking EM radiation from a complex source; 
extending this to other wave problems such as in vibro-acoustics
or quantum  mechanics is straightforward. The phase-space 
representation based on the Wigner function
provides a useful means of physically interpreting the propagated
data. It also serves as a very efficient computational technique both for an
exact propagation of CFs and in terms of a 
ray approximation leading to the Frobenius-Perron operator. 
This provides a good description of the propagated data even 
when applied to source data that are relatively far from
homogeneity. Where necessary, more heterogeneous sources can be 
accounted for by higher-order approximations leading to an Airy propagator.
This propagator proved important in the case of a planar random source emitting 
in presence of a planar reflector, for which we are able to reconstruct the fine structure 
of the phase space in presence of interference. Evanescent decay into the near field can 
also be accounted for using simple propagation rules. These rules has been used to investigate 
the effect of evanescent waves in near-field correlation 
functions. For source correlations exhibiting smaller-than-wavelength scales, we predicted a rapid 
initial increase of the correlation length (with distance from the source), before  
it saturates with the onset of the Van Cittert-Zernike behaviour at a distance of a wavelength.
The approximations used have been validated  through full-wave simulations using model sources and 
numerical sources exhibiting strong statistical inhomogeneities. 

\section{Acknowledgments}
{
Financial supported through by the EPSRC  (Grant-Ref.: EP/K019694/1) is gratefully acknowledged. 
}

\bibliography{Wigner_biblio}
\end{document}